%% file: SchrodNewton_New3.tex
\documentclass[11pt,fleqn]{article}
\usepackage{amsmath,graphicx}

\newcommand{\be}{\beq\label}
\newcommand{\ee}{\eeq}
\newcommand{\p}{\prime}

\input preamble.tex

\begin{document}

\Title{Relativistic generalization of the  
Schr{\"o}dinger-Newton model \\ for the wavefunction reduction }

\Aunames{Vladimir V. Kassandrov\auth{a,1} and Nina V. Markova\auth{b,2}}

\Addresses{
\addr a {Institute of Gravitation and Cosmology, Peoples' Friendship
	University of Russia (RUDN University), 6 Miklukho-Maklaya St., Moscow, 117198, Russian Federation}
\addr b {S.M. Nikolsky Mathematical Institute, Peoples' Friendship
	University of Russia, (RUDN University), 6 Miklukho-Maklaya St., Moscow, 117198, Russian Federation}
	}

\Abstract
  {We consider the model of the self-gravity driven spontaneous wavefunction reduction proposed by L. Di\'osi, R. Penrose et al. and based on a self-consistent system of the  Schr{\"o}dinger and Poisson equations. An analogous system of coupled Dirac and Maxwell-like equations is proposed as a relativization. Regular solutions to the latter form a discrete spectrum in which all the ``active'' gravitational masses are always positive, and approximately equal to inertial masses and to  the mass $m$ of the quanta of Dirac field up to the corrections of order $\alpha^2$. Here $\alpha=(m/M_{pl})^2$ is the gravitational analogue of the fine structure constant negligibly small for nucleons. In the limit $\alpha \to 0$ the model reduces back to the nonrelativistic Schr{\"o}dinger-Newton one. The equivalence principle is fulfilled with an extremely high precision. The above solutions correspond to various states of the same (free) particle rather than to different particles. These states possess a negligibly small difference in characteristics but essentially differ in the widths of the wavefunctions.  For the ground state the latter is $\alpha$ times larger the Compton length, so that a nucleon cannot be sufficiently localized to model the reduction process}




\email 1 {vkassan@sci.pfu.edu.ru}
\email 2 {n.markova@mail.ru}

\section{Wavefunction reduction and the Schr{\"o}dinger-Newton model}	

In a number of works\cite{Karolyhazy,Diosi,Penrose96} 
it has been argued that the phenomenon of wavefunction reduction is an {\it objective} process driven by the gravitational self-interaction between the parts of a (distributed) quantum particle -- the ``lumps'' of probability.

Such a reduction occurs from a superposition state of a free quantum particle to one of its basic ``eigenstates''. To ensure random outcomes of measurements and, in particular, the Bohr's rule for probabilities, a number of {\it stochastic} wavefunction collapse models has been suggested~\cite{Girardi,Bassi,Adler}.

The ``eugenstates'', in the nonrelativistic approximation, correspond to regular stationary solutions of the self-consistent PDE system for coupled  {\it Schr{\"o}dinger and Poisson equations},
\begin{equation}\label{SNP}
-\frac{\hbar^2}{2m}\Delta \psi +U\psi = E\psi, ~~\Delta U = 4\pi Gm^2 \vert \psi\vert^2 \, ,  
 \end{equation}
$G$ being the Newton's gravitational constant.
 
The above {\it Schr{\"o}dinger-Newton} model (hereafter the SN model)~\footnote{Possible fundamental status and derivation of the SN equations were discussed in a large number of works (see, e.g., Refs.\cite{Bahrami,Hu,Guilini})}  is effectively nonlinear: 
the gravitational self-energy $U$ which defines $\psi$-function is itself determined by the probability distribution $\vert \psi \vert^2$.

Solutions to (\ref{SNP}) can be found only numerically. In Refs.\cite{Bernstein, 
Moroz}, in the spherically symmetric case, it was shown that 
the spectrum of {\it everywhere regular and normalized} ``eigenstates''  is {\it discrete} and can be described by an integer $n=N+1=1,2,\cdots$ related to the number of {\it nodes} $N$  of the wavefunction $\psi(r)$. 
At great distances $\psi$ decreases exponentially while $U\approx -m/r$ is asymptotically Newtonian.

For large $n$, the energy eigenvalues satisfy the hydrogen-like dependence
\begin{equation}\label{Bohr}
E_n = - E_0/n^\gamma, ~~~\gamma \approx 2 \, , 
\end{equation}
where, say, for a nucleon $E_0$ is about $10^{-76}$ of its rest energy.

However, the mass $m$ in Eqs.~(\ref{SNP}) is an arbitrary external parameter, and, say, the gravitational field in the SN model does not at all contribute to its value.  To obtain the rest mass (or, better, a whole mass spectrum) and spin,  one can (as in the canonical QM) to replace Schr{\"o}dinger's equation by the Dirac one. 

As to the relativistic counterpart of ``Newtonian gravity'', the {\it self-consistent} Dirac-Einstein system seems too complicated. As a first step, one can instead consider\cite{Newman,Kibble}  the so-called {\it Maxwellian gravity}. We shall see below that the latter describes the structure and gravitational self-interaction of a quantum particle in quite a remarkable way. 

\section{`Maxwellian gravity' and self-consistent Dirac-Maxwell model}           

We come thus to a field system represented by the Lagrangian
\begin{equation}\label{DGM}
\begin{array}{lll}
L=-\frac{1}{16\pi}F^2 + \frac {\hbar c}{2}(i\bar \psi \gamma^\mu \partial_\mu \psi - i \partial_\mu \bar \psi \gamma^\mu \psi) \\ 
\\
+mc^2 \bar \psi \psi + e A_\mu \bar \psi \gamma^\mu \psi \, , 
\end{array}
\end{equation}
where $F^2:=F_{\mu\nu}F^{\mu\nu},~F_{\mu\nu}=\partial_\mu A_\nu - \partial_\nu A_\mu$ is the Maxwell invariant, and $e:=m\sqrt G$ is the gravitational charge -- ``active'' mass of a particle. Remarkably, $L$ is actually a {\it classical gravitational analogue} of the operator equations of quantum electrodynamics.  

And indeed, in the framework of {\it classical spinor electrodynamics}  the above model, in which the coupling constant $e$ corresponds now to the elementary electric charge,  has been repeatedly proposed\cite{Wakano,Kassandrov,KassTerl,Bohum} as a {\it unified field model} for charged fermions.

However, in all of the above cited works the {\it rest masses (energies) of the  particlelike solutions take negative values} despite the positive definite contribution of the Maxwell term $F^2$.
Consequently, two sufficiently separated like charges-solitons attract each other, and this is a principal difficulty of the Dirac-Maxwell model for charged fermions. However, this becomes an advantage in the gravitational framework of the Dirac-graviMaxwell model (\ref{DGM}) (hereafter the DGM model).  

\section{Reduction of  the Dirac-gravi-Maxwell to the Schr{\"o}dinger-Newton system and  the equivalence principle}

Let us make an appropriate {\it scale transformation} 
\be{scale}
x_\mu \mapsto  x_\mu/k, ~~\psi \mapsto \sqrt{k^3/4\pi \alpha}~ \psi,~~ A_\mu \mapsto (kG\sqrt{m}/\alpha) A_\mu, 
\ee
where $k:=mc/\hbar$ is the inverse Compton lemgth and $\alpha=(m/M_{Pl})^2 = Gm^2/\hbar c$ is the gravitational analogue of the fine structure constant. Then   
the DGM Lagrangian  transforms to a completely {\it dimensionless} form, for which the field equations read 
 \begin{equation}\label{eqs}
 \begin{array}{ll}
 i\gamma^\mu (\partial_\mu - iA_\mu)\psi + \psi = 0, ~~\Box A^\mu =\bar \psi \gamma^\mu \psi, \\ 
 (\partial_\mu A^\mu =0)\, .
 \end{array}
 \end{equation}
The only {\it dimensionless} parameter enters only the {\it normalization condition}
\begin{equation}\label{norm}
 \frac{1}{4\pi} \int \psi^+ \psi~ dV =\alpha.
 \end{equation}
  
In the stationary case $\Delta \Phi = \psi^+\psi$ and, if  condition (\ref{norm}) is satisfied, the  gravitational charge ($\sim$ active mass) is always the same, equal in modulus to $e=m\sqrt G$.  As to the (dimensional) rest energy $W$ ($\sim$ inertial mass) of the particle, in Refs.~\cite{Kassandrov,KassTerl} the following simple expression has been obtained:
\begin{equation}\label{energy}
W= - \frac{mc^2}{\alpha} \int \bar \psi \psi~ dV \, . 
 \end{equation}

The DGM system (\ref{DGM}) does not possess spherically symmetric solutions \cite{Wakano,Hydrogen}. We restrict by the following principal class of stationary axisymmetric solutions:
\begin{equation}\label{axial}
\psi^T =(\kappa, \chi) e^{-i\omega t},~~ A_\mu = \{\Phi, \vec A\} \,,
\end{equation}
where for the 2-spinors and gravi-potentials one takes, respectively:
\begin{equation}\label{2spinors}
\begin{array}{ll}
\kappa^T = a(r) (1~0),~~\chi^T =ib(r) (\cos {\theta},~ \sin{\theta}~ e^{i\varphi})\, , \\
\Phi = \Phi(r), ~~A_\varphi = \Lambda(r) \sin {\theta}, ~(A_r = A_\theta =0) \,.
\end{array}
 \end{equation}

Now, to make use of the small value of $\alpha$, one transforms  the field functions and radial coordinate as follows:
\begin{equation}\label{fieldeq}
\begin{array}{lll}
a =\alpha^2 (1-\omega)^{3/2} A, ~~b=\alpha^3 (1-\omega)^{3/2} B \, , \\
\Phi = \alpha^2 (1-\omega) \phi, ~~\Lambda=\alpha^3 (1-\omega) \lambda \,, \\
r=\rho/\alpha (1-\omega). 
\end{array}
 \end{equation}
 
Neglecting then terms of order $\alpha^2$ in the equations for spinor functions (which correspond to gravimagnetic corrections) and for gravitational potential $\phi$, one reduces the field equations to 
 \begin{equation}\label{nonrel}
  \begin{array}{lll}
 A^\prime = - B, ~~B^\prime +\frac{2}{\rho} B = (E - \phi)A \, ,\\
 \phi^{\prime\prime} + \frac{2}{\rho}\phi^\prime = A^2, \\
 E:=- (1 +\omega) / \alpha^2 (1-\omega)<0 \, , 
 \end{array}
 \end{equation}
which, after substitution of $B$ into the second Eq.~(\ref{nonrel}),  {\it takes the form exactly equivalent to the non-relativistic SN system} (\ref{SNP}) (in the spherically symmetric case)
\be{SNspher}
\begin{array}{lll}
A^{\p\p} +\frac{2}{\rho} A^\p = (\phi-E)A \, ,\\
 \phi^{\p\p} + \frac{2}{\rho}\phi^\p = A^2, \\
 \int A^2 \rho^2 d\rho =1.
 \end{array}
\ee
 As for gravimagnetic potential $\lambda(\rho)$, it can be computed for an already obtained solution of (\ref{nonrel})  from the equation
\be{magn}
 \lambda^{\p\p}  + \frac{2}{\rho}\lambda^\p - \frac{2}{\rho^2} \lambda = 2AB.
\ee

Thus, under the requirement of regularity and normalization of $\psi$-function, one again obtains a discrete spectrum of eigenstates and eigenvalues of the {\it binding energy} $E$ as in the nonrelativistic case. Now, however, {\it the exact expressions for the total rest energy} $W$ (inertial mass) and for the gravitational charge ($\sim$ active mass) $Q$ in the dimensional form read
\begin{equation}\label{restenergy}
\begin{array}{cc}
\vert W \vert=mc^2 \int (A^2 - \alpha^2 B^2) \rho^2 d\rho \, , \\ 
\vert Q \vert = m{\sqrt G} \int (A^2 + \alpha^2 B^2) \rho^2 d\rho \, . 
\end{array}
\end{equation}
Since, say, for a nucleon the ``gravitational fine structure constant'' $\alpha=(m/M_{Pl})^2$ is negligibly small ($\alpha^2 \sim 10^{-76})$ one concludes the coincidence of active gravitational and inertial masses with extremely high precision, that is, the {\it fulfillment of the equivalence principle for elementary particles}. Moreover,  in the limit $\alpha \to 0$, the rest energy and gravitational  charge are proportional to the same normalization integral $\int A^2 \rho^2 d\rho =1$, so that for any eigenstate one gets the same universal values, 
\begin{equation}\label{unvrsl}
\vert W \vert \approx mc^2, ~~\vert Q \vert \approx m{\sqrt G}, 
\end{equation}
and the total angular momentum $J=\hbar/2$, Refs.~\cite{Kassandrov,KassTerl}.

\section{Conclusion. Position eigenstates and the reduction process}

The effectively nonlinear DGM system (\ref{DGM}) has been considered and used for description of fermions bound by their own gravitational field. In the stationary axisymmetrical case, everywhere regular and normalized solutions form discrete spectrum of ``eigenstates''  and corresponding ``eigenvalues'' -- Noether charges. 

In the nonrelativistic limit $\alpha \to0$ the DGM system reduces to the SN system. However, {\it a  new class of solutions has been found}~\footnote{It will be presented in a forthcoming paper.} (which correspond to the exchange $\kappa \leftrightarrow \chi$ in Eq.~(\ref{2spinors})), and gravimagnetic corrections (of order $\alpha^2$) calculated. 

For all the above solutions the active gravitational mass is proportional to the inertial one and equal to the mass scale factor $m$ in the Dirac equation. Thus, the equivalence principle holds up to (negligibly small) corrections of order $\alpha^2$. At far distances fermions attract each other $\sim 1/r^2$. From (\ref{unvrsl}), one also concludes that  
different solutions do not correspond to different particles, rather to different ``eigenstates'' of the same particle. 

Note finally that in both the SN and DGM models minimal ``spread'' of the wavefunction  
 is $\alpha = (m/M_{Pl})^2\equiv Gm^2/hc$  times larger the Compton length $L=\hbar/mc$.  For the DGM model, this follows from the successive scale transformations of coordinates given by (\ref{scale}) and (\ref{fieldeq}). As for the nonlelativistic SN system of equations (\ref{SNP}), it is {\it scale invariant by itself}. However, together with the normalization condition, it can be reduced to the proper dimensionless form (\ref{SNspher}) only via the uniquely defined transformation of the form 
\be{scaleSN}
 r \mapsto ( \hbar^2 / Gm^3)r \equiv (L / \alpha)r, ~~U \mapsto mc^2 \alpha^2 U, ~~\psi \mapsto L^{-3/2}\psi. 
 \ee
 Therefore, a nucleon cannot be essentially localized, and description of the reduction process itself is in fact problematic in the framework of both the SN and DGM models.

\section*{Acknowledgments}

The authors are grateful to J. A. Rizcallah for valuable comments. 
The publication has been prepared with the support of the ``RUDN University Program 5-100''.

\end{document}

%% file: preamble.tex
\usepackage{amsfonts,amssymb,cite}
\usepackage{epsf,graphicx}

\def\noi{\noindent}

\newcommand{\Title}[1]{\noi {{\Large\bf #1}}\\[1ex]}

\def\Aunames#1{\noi{\bf #1}}
\def\auth#1{${}^{#1}$}
\def\Addresses#1{\medskip\noi \protect
	\begin{description}\itemsep -3pt {\it #1} \end{description}}
\def\addr#1#2{\item[${}^{#1}$]{\it #2}}

\newcommand{\Abstract}[1]{\vskip 2mm \begin{center}
        \parbox{16.4cm}{\small\noi #1} \end{center}\medskip}
\newcommand{\PACS}[1]{\begin{center}{\small PACS: #1}\end{center}}

\def\email#1#2{\footnotetext[#1]{e-mail: #2}\addtocounter{footnote}{1}}

\def\nqq{\hspace*{-2em}}

\def\Jl#1#2{#1 {\bf #2},\ }

\def\ApJ#1 {\Jl{Astroph. J.}{#1}}
\def\CQG#1 {\Jl{Class. Quantum Grav.}{#1}}
\def\DAN#1 {\Jl{Dokl. AN SSSR}{#1}}
\def\GC#1 {\Jl{Grav. Cosmol.}{#1}}
\def\GRG#1 {\Jl{Gen. Rel. Grav.}{#1}}
\def\JETF#1 {\Jl{Zh. Eksp. Teor. Fiz.}{#1}}
\def\JETP#1 {\Jl{Sov. Phys. JETP}{#1}}
\def\JHEP#1 {\Jl{JHEP}{#1}}
\def\JMP#1 {\Jl{J. Math. Phys.}{#1}}
\def\NPB#1 {\Jl{Nucl. Phys. B}{#1}}
\def\NP#1 {\Jl{Nucl. Phys.}{#1}}
\def\PLA#1 {\Jl{Phys. Lett. A}{#1}}
\def\PLB#1 {\Jl{Phys. Lett. B}{#1}}
\def\PRD#1 {\Jl{Phys. Rev. D}{#1}}
\def\PRL#1 {\Jl{Phys. Rev. Lett.}{#1}}

\def\lal{&&\nqq {}}

\def\beq{\begin{equation}}
\def\eeq{\end{equation}}
\def\bear{\begin{eqnarray}}
\def\bearr{\begin{eqnarray} \lal}
\def\ear{\end{eqnarray}}
\def\earn{\nonumber \end{eqnarray}}

